# PSpice modelling of the Pulsed Electro Acoustic method for dispersive polymer sample application

**Aurélien Pujol[1], Laurent Berquez[1], Fulbert Baudoin[1], Denis Payan[2]**

1. Laboratoire plasma et conversion d'énergie, UMR 5213, CNRS-INP-UPS, 31062 Toulouse, France

2. Centre national d'études spatiales, 18 avenue Edouard Belin, 31401 Toulouse cedex 9, France

## Abstract

For space application, space charge inside dielectric materials can lead to unwanted phenomena as breakdown and electrostatic discharge (ESD)[1]. To prevent these incidents, the Pulsed Electro Acoustic device is employed to measure the space charge of dielectric samples. Unfortunately, most of the current PEA signal processing are inaccurate to study dielectric material that attenuate and disperse the acoustic wave, as well as study multi-layer or thin samples. A model under the PSpice software is developed in this work with the aim of improving the PEA signal processing. This model is an evolution of previous works including the attenuation and dispersion of a polymer sample using the Nearly Local Kramers-Kronig (NLKK) relation. This model of polymer sample is implemented into PSpice as a frequency dependent transmission line (FDTL) using the Branin's Method of Characteristics (MoC) model. A comparison of a model simulation and an experimental PEA signal gotten from a PTFE sample is presented.

## Introduction

The space industry has studied the satellite space charge for understanding and anticipating such phenomena as electrostatic discharge (ESD)[2–4]. One of the most used methods to study space charge inside a dielectric insulator is the Pulsed Electro Acoustic (PEA) introduced by the research group of Takada *et al.*[5–10]. It is a non-destructive method for exciting the charges by an electrical stress and measuring the acoustic response with a piezoelectric sensor. The PEA method requires a mathematical post-processing to recover accurate charges distribution from measurement. This processing involves a calibration step, using an experimental calibration signal, to define the PEA impulse response from well-known sheet capacitive charges at the interface between the dielectric sample and the ground electrode[7]. Unfortunately, the information from the only ground electrode does not take into account the acoustic wave attenuation and dispersion as well as reflections near the semi-conducting High Voltage (HV) electrode or inside multi-layer sample. Moreover, the current calibration step cannot be employed for thin sample (under 50 microns) because the impulse response from the ground electrode is indistinguishable from the impulse response from the HV electrode. In this study, a PEA PSpice model taking into account the acoustic wave attenuation and dispersion are introduced. The final aim is to use this model in future studies to improve the current calibration step by generating the PEA responses at different space positions inside the sample.

Bernstein[11] proposed an analysis of the whole PEA device. He developed an analytical model to describe the acoustic force generation and its propagation through the PEA cell. He also suggested a model of the piezoelectric sensor to convert an input acoustic wave to an output voltage one. His work led to some studies that implemented his model using the COMSOL software in order to understand the impact of the PEA cells configuration on the acoustic wave[12,13]. However, these models do not take into account the dielectric losses of the piezoelectric sensor and the electrical acquisition chain as amplifiers and oscilloscope. These lacks motivated the development of PEA electrical models using the MATLAB[14] software or the PSpice software[15–17]. The sensor is based on the electrical Redwood's model[18] and the acoustic propagation media are modeled by transmission lines (TL) according to an acoustic-electric analogy. Using this model, Galloy[15] studied the impact of the dielectric losses of the piezoelectric and Chahal[17] highlighted the impact of the amplifier presence.

However, acoustic dispersion of the dielectric sample has not been processed yet for transient analysis. Chahal[16] already took into account acoustic attenuation of PEA materials as constants which didn't lead to any dispersion. Furthermore, Püttmer[19], Dahiya[20], Galloy[15], and Chahal[17] considered frequency dependency for the losses but not for the acoustic velocity of the material, which yielded no dispersion and didn't respect the material causality too. Moreover, the main current methods used to measure the dielectric sample losses do not consider the effect of the High Voltage electrode impedance.

In this work, we present a PSpice model based on previous studies of the authors cited above but including attenuation and dispersion. First, we modify the current acoustic-electric analogy to consider an acoustic medium that attenuate and disperse as an electrical transmission line that attenuate and disperse. Second, we present a tool to implement this frequency dependent transmission line (FDTL) into the PSpice software. Then, we show the effect of the HV electrode impedance on the output PEA signal. Finally, the simulation of the new PEA model is presented, compared, then validated with an experimental PEA measurement.

## I. Electroacoustic modelling of the PEA

### 1. PEA modelling with PSpice

The PEA is set under biasing configuration as shown in FIG. 1. A DC voltage generator $V_{DC}$ creates capacitive charges $\sigma_c$ at each electrode. In parallel there is a pulse generator to create an exciting electrical field $E_P$. The dipoles $R$ and $C$ are used to decouple the generators. The exciting head is integrating the connections between the both generators and an aluminum electrode, which is affixed to the semi-conducting HV electrode. The role of the HV electrode is to ensure good acoustic contact and reduce the reflections between the aluminum electrode and the sample. The clamping screw is the bottom of the PEA cell, which is used to well compress the sample to the both HV and ground electrodes.

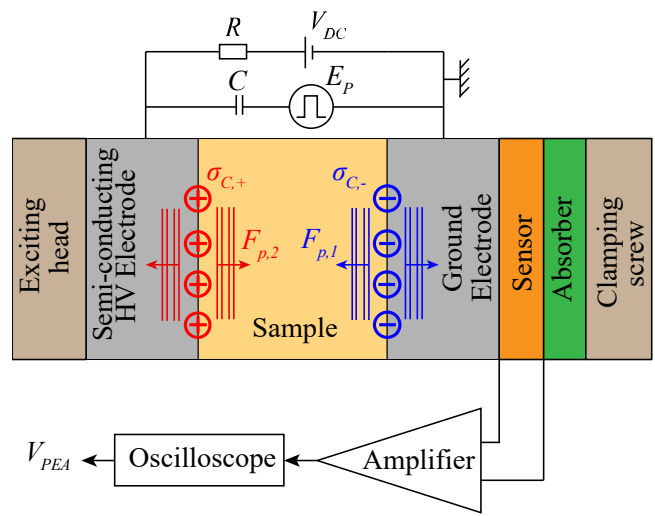

*FIG. 1. Block diagram of the PEA cell*

For a sample that is no polar, no piezoelectric and if we neglect the electrostriction, $F_{p,1}$ is the acoustic force created by the convolution between the electrical pulse $E_P$ and the capacitive charges $\sigma_{C,-}$ at the ground electrode interface[21]. $F_{P,2}$ is the one generated from the semi-conducting HV electrode. For further details, the PEA device is well described in the review of Imburgia[22].

Respecting the acoustic-electric analogy these forces are modeled by voltage generators in the PSpice model shown in FIG. 2. The acoustic media are modeled by transmission lines.

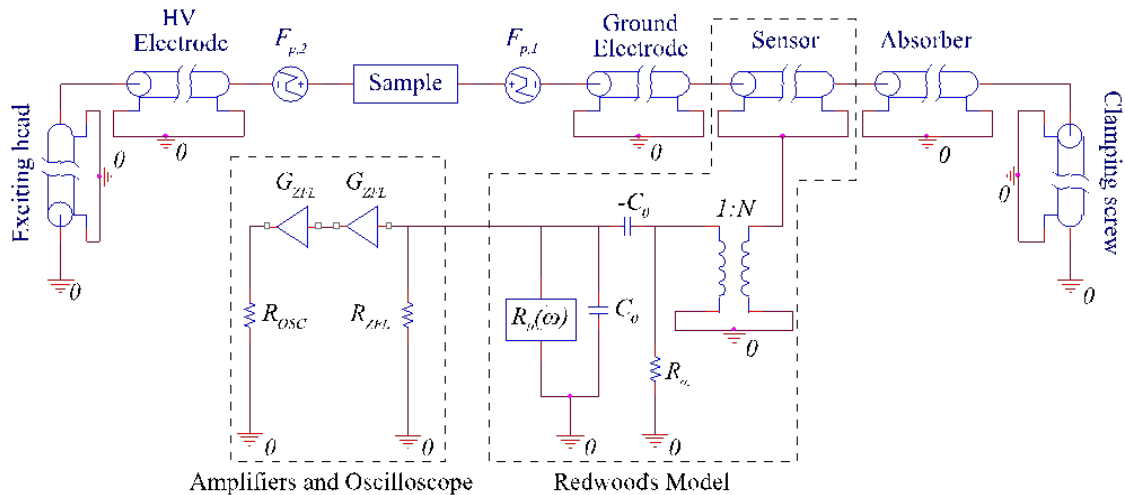

FIG. 2. PSpice model of the PEA

Note that the acoustic forces are placed at the interfaces between the polymer and the electrodes although the capacitive charges, so the resultant forces, are theoretically inside the polymer. This approximation is made because the capacitive charges penetration (less than 1 micron) is very low compared to the PEA space resolution. The acoustic sensor is implemented using the Redwood's Model[18] which is based on the Masson's one[23]. The acoustic-electric conversion is made by an ideal transformer (with a transformation ratio $N$) coupled with the clamped capacity $C_0$ and the flexibility $-C_0$. The dielectric losses of the sensor are modeled by a frequency dependent resistance $R_0(\omega)$ implemented into PSpice using the approach of Dahiya[20]. For the measurement chain, each of the parallel ZFL-500LN+ amplifiers used is modeled by an input $50\ \Omega$ resistance $R_{ZFL}$ and by a $28\ dB$ voltage gains $G_{ZFL}$. A $50\ \Omega$ input resistance $R_{osc}$ is to model the input impedance of the oscilloscope. Ideal transmission lines are used to represent the acoustic wave propagation inside every acoustic media (cf. Equations (1)-(2)), excepting the polymer sample. Indeed, in this study the attenuation and the dispersion of the sample are considered to get a better simulation of the acoustic wave coming far from the sensor. The following part reminds the basics of the acoustic-electric analogy to understand how the acoustic frequency dependent losses and velocity are modeled in a transmission line.

## 2. Acoustic-electric wave propagation analogy for polymer that attenuate and disperse

The analogy was detailed and used by Deventer[24] and Wei[25]. Equations (1) to (4) shows how to model an acoustic medium by a transmission line. $L$ is the linear inductance which represents the linear mass of the medium. This linear mass depends on the density $\rho$ and the surface $S$. The linear capacitance $C$ corresponds to the elasticity of the medium which is determined by the linear mass and the sound phase velocity $c_{ac}$. The linear resistance $R$ leans on the viscous loss factor $\alpha_v$. The linear conductance $G$ expresses the thermal loss factor $\alpha_{th}$.

$$L = \rho S \tag{1}$$

$$C = \frac{1}{\rho S c_{ac}^2} \tag{2}$$

$$R = 2\rho S c_{ac} \alpha_v \tag{3}$$

$$G = \frac{2}{\rho S c_{ac}} \alpha_{th} \tag{4}$$

These relations are valid if the TL losses are low enough, which means if for a given pulsation $\omega$, $R \ll L\omega$ and $G \ll C\omega$.

Equations (1) and (2) are used to model all the media except the sample. Equations (1) to (4) are used to model the sample. Note that for polymer material, the thermal losses are negligible compared to the viscous losses and leads to consider the conductance nil ($G \cong 0$). Some studies showed that the polymer viscous losses are not constant and depend on the frequency[26–29]. Assuming that hysteresis absorption is the dominant loss process, the polymer attenuation coefficient $\alpha_v(f)$ could be modeled by a linear function of the frequency:

$$\alpha_v(f) = a_0 + a_1 f \tag{5}$$

The zero frequency component $a_0$ is attributed to the end of the tail relaxation[26]. To respect the causality of the material, the Nearly Local Kramers-Kronig (NLKK) relation links $c_{ac}(f)$ and $\alpha_v(f)$[30–33]:

$$\frac{\pi^2 f^2}{c_{ac}^2(f)} \frac{dc_{ac}(f)}{df} \cong \alpha_v(f) \tag{6}$$

Integration of (6) from a starting frequency $f_0$ yields:

$$\frac{1}{c_{ac}(f_0)} - \frac{1}{c_{ac}(f)} \cong \frac{1}{\pi^2} \int_{\phi=f_0}^{f} \frac{\alpha_v(\phi)}{\phi^2} d\phi \tag{7}$$

If, for the frequency of interest,

$$|c_{ac}(f) - c_{ac}(f_0)| \ll c_{ac}(f_0) \tag{8}$$

then the left-hand side of (7) is approximately equal to $[c_{ac}(f) - c_{ac}(f_0)]/c_{ac}^2(f_0)$. The NLKK relation yields:

$$c_{ac}(f) = c_{ac}(f_0) + \left[\frac{c_{ac}(f_0)}{\pi}\right]^2 \int_{\phi=f_0}^{f} \frac{\alpha_v(\phi)}{\phi^2} d\phi \tag{9}$$

Integration of (5) in (9) yields[28,29,33]:

$$c_{ac}(f) = c_{ac}(f_0) + \left(\frac{c_{ac}(f_0)}{\pi}\right)^2 \left[a_1 ln\left(\frac{f}{f_0}\right) + a_0 \left(\frac{1}{f} - \frac{1}{f_0}\right)\right] \tag{10}$$

This relation is valid for materials having no sharp resonances in the frequency range of interest, and if[32,33] :

$$\left[\frac{\alpha_v(f)c_{ac}(f)}{2\pi f}\right]^2 \ll 1 \qquad (11)$$

Substitution of (5) and (10) into (11) yields:

$$\left[\frac{(a_0 + a_1 f)}{2\pi}\right]\left[\left[c_{ac}(f_0) + \left(\frac{c_{ac}(f_0)}{\pi}\right)^2\left[a_1 ln\left(\frac{f}{f_0}\right) + a_0\left(\frac{1}{f} - \frac{1}{f_0}\right)\right]\right]\right]^2 \ll 1 \qquad (12)$$

It appears that this condition is right if $a_0 = 0$. Under this last assumption, the material is causal and the NLKK relation imposes the next relations for the viscous losses and the acoustic velocity:

$$\alpha_v(f) = a_1 f \qquad (13)$$

$$\Rightarrow \quad c_{ac}(f) = c_{ac}(f_0) + \left(\frac{c_{ac}(f_0)}{\pi}\right)^2 \left[a_1 ln\left(\frac{f}{f_0}\right)\right] \qquad (14)$$

Equations (13) and (14) are substituted into (3) and (2) to model a polymer material that attenuate and disperse acoustic waves. Unfortunately, the PSpice lossy transmission line component does not allow to add frequency dependent passive elements such as a capacitance. A common tool for frequency dependent transmission line is then explained in the next part.

Note that Galloy[15] and Püttmer[19] proposed a transmission line taking into account frequency dependent polymer losses using the mechanical loss angle $\delta_m$ in the resistance $R(\omega) = \rho S \delta_m \omega$. While Dahiya[20] and Chahal[17] use the mechanical loss angle in the conductance $G(\omega)$. However they consider a constant acoustic velocity and in that case the polymer causality is not respected for transient analysis[34].

### 3. Implementation of a FDTL into the PSpice model

To implement a frequency dependent transmission line into PSpice, the Branin's Method of Characteristics (MoC) model[35] is employed for the generation of line macro model suitable for transient analysis. The transmission line is treated as a multiport, where $[V_1(s)\,;\,I_1(s)]$ are the near end input and $[V_2(s)\,;\,I_2(s)]$ are the far end output voltage-current port quantities in the Laplace ($s = j\omega$) domain. For a line with a characteristic impedance $Z_{el}(s)$, a complex propagation vector $\gamma(s)$, and a length $l$, the solutions of the Telegrapher's equations in the Laplace domain can be reduced to the following[36,37]:

$$V_1 = Z_{el}(s)I_1 + (V_2 + Z_{el}(s)I_2)e^{-\gamma(s)l} \qquad (15)$$
$$V_2 = Z_{el}(s)I_2 + (V_1 + Z_{el}(s)I_1)e^{-\gamma(s)l} \qquad (16)$$

These equations are equivalent to the electrical network shown below:

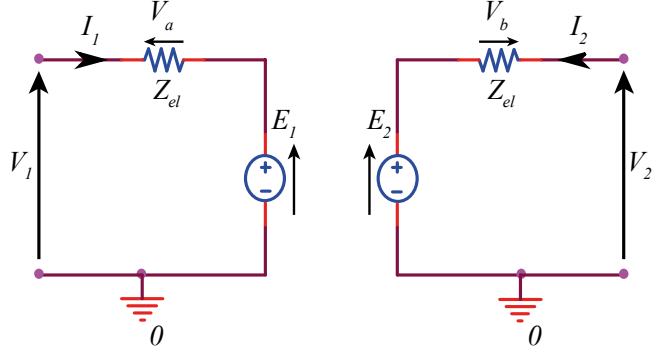

FIG. 3. Thevenin equivalent circuit for the Branin's Method of Characteristics model of a FDTL

With:

$$E_1 = (V_2 + V_b)e^{-\gamma(s)l} \quad (17) \qquad E_2 = (V_1 + V_a)e^{-\gamma(s)l} \quad (18)$$

$$V_a = Z_{el}(s)I_1 \quad (19) \qquad V_b = Z_{el}(s)I_2 \quad (20)$$

$E_1$ and $E_2$ are Laplace expressions implemented in PSpice using the controlled voltage source "ELAPLACE"[38]. This above circuit is implemented into PSpice.

The complex wave vector $\gamma$ and the characteristic impedance $Z_{el}$ are defined using the electric-acoustic analogy[24,25]:

$$\gamma(s) \cong \frac{R(s)}{2}\sqrt{\frac{C(s)}{L}} + s\sqrt{LC(s)} = \alpha_v(s) + \frac{s}{c_{ac}(s)} \quad (21)$$

$$Z_{el}(s) \cong \sqrt{\frac{L}{C(s)}} = \rho S c_{ac}(s) \cong \rho S c_{ac}(j2\pi f_0) \quad (22)$$

$Z_{el}(s)$ is approximate as a constant real number and is implemented into PSpice using a casual resistance. Indeed, its reactance part can be neglected because it only causes a low phase shift that is negligible compared to the wave dispersion. These equations give a straight way to define a FDTL in PSpice using the acoustic parameters $\rho, S, \alpha_v$ and $c_{ac}$ of a polymer material. The complex wave vector is an irrational function of the complex frequency $s$. It has to respect the "TL-causality" requirement to insure that any excitation signal entering one end of the transmission line will appear at the other end only after the time-of-flight delays[37]. The causality requirement is respected because of the NLKK relation which defines the complex wave vector does[37,39]. Furthermore, an advantage of this FDTL model is that the causality requirement can be guaranteed by enforcing the output at the far end of the line to be zero before the time of flight. It can be done by adding a pure delay $T$ at the propagation function $\gamma$.

4. Validation of the polymer material FTDL model

To check the FDTL model, the first is to set it with the following parameters of an XLPE polymer sample, $l = 1\ mm$, $S = 79\ mm^2$, $\rho = 930\ kg/m^3$. The parameters used to set the attenuation factor and the velocity (cf. Equations (13) and (14)) of the XLPE sample are taken from another work about cable PEA[29].

$$a_1 = 6{,}2.\,10^{-5}\ (m.Hz)^{-1}$$

$$f_0 = 2\ MHz$$

$$c_{ac}(f_0) = 1900\ m/s$$

Then, the FDTL model is added into the following PSpice network:

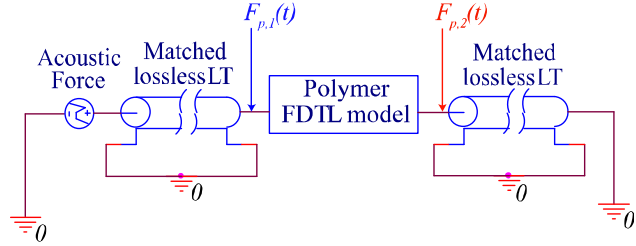

FIG. 4. Schematic of the PSpice network to check the polymer FDTL model

The idea is to set an acoustic force upstream of the polymer FDTL, then probe both output forces $F_{p,1}$ and $F_{p,2}$ from each end of the FDTL. $F_{p,2}$ is the one that crossed it and $F_{p,1}$ is the one that did not. Some matched lossless lines are added to include delays and avoid any reflection in the probed signal. The input acoustic force chosen is a 5 $ns$ width theoretical square pulse with 1 $N$ amplitude. Note that a frequency analysis with a sine wave input could have been used too. The plot of the forces around each end of the FDTL model are shown below:

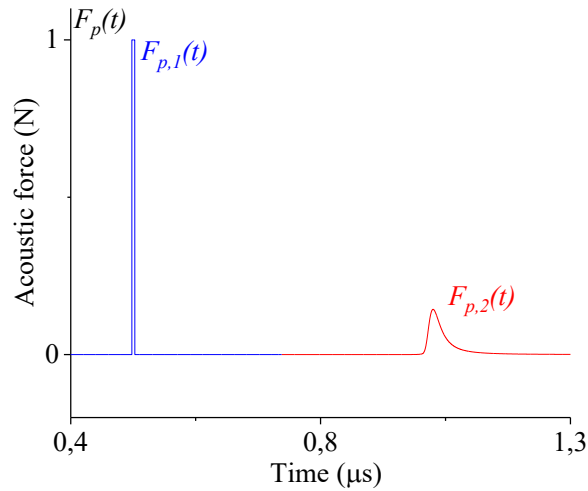

FIG. 5. Acoustic forces at each end of the FDTL

$F_{p,2}(t)$ is the wave propagation solution of the input $F_{p,1}(t)$ at the end of the line. In the frequency domain, these forces are related as follows:

$$F_{p,2}(j\omega) = F_{p,1}(j\omega)e^{-\gamma(f)l} = F_{p,1}(j\omega)e^{-\left[\alpha_v(f)+\frac{j\omega}{c_{ac}(f)}\right]l} \quad (23)$$

From this expression, the attenuation factor and the phase velocity can be calculated from the Fast Fourier transform (FFT) of the two transient probed components $F_{p,1}(t)$ and $F_{p,2}(t)$. Ditchi et al.[40] first suggested this procedure to determine the acoustic attenuation and phase velocity in solid insulating material:

$$\alpha_v(f) = -\frac{1}{l} \ln\left(\left|\frac{FFT(F_{p,2}(t))}{FFT(F_{p,1}(t))}\right|\right) \tag{24}$$

$$c_{ac}(f) = \frac{-2\pi f l}{\varphi\left(FFT\left(F_{p,2}(t)\right)\right) - \varphi\left(FFT\left(F_{p,1}(t)\right)\right)} \tag{25}$$

These parameters found by applying Equations (24) and (25) on the simulated signal of the FDTL model are then compared to the NLKK theoretical ones given by Equations (13) and (14) using with XLPE polymer acoustic parameters:

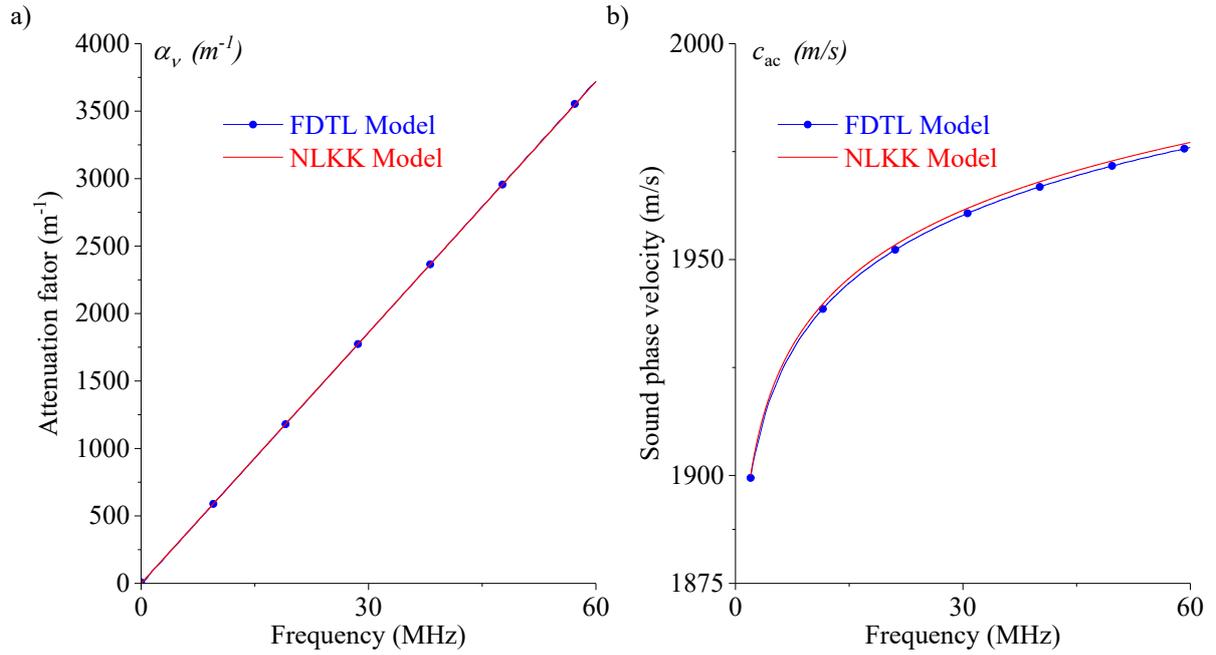

FIG. 6. (a) Attenuation factor, (b) Wave velocity

The bandwidth of 60 $MHz$ is representative of our PEA measurements. The percentage ratio between the norm of the difference of the attenuation models and the norm of the attenuation NLKK model is 2.2 %. The same ratio is 0.2 % for the wave velocity. These ratios are satisfactory to validate the FDTL model of a dispersive polymer material.

### 5. Experimental precaution

Note that in a real experimental situation, the polymer material is not surrounded by matched TL but is placed between unmatched HV and ground electrodes. FIG. 7 shows the PSpice network of the PEA under biasing configuration which means that both acoustic forces around the sample have the same amplitude $F_a$ but with opposite signs. This time one probe is used to plot $F_p$ which is the sum of the acoustic force $F_{p,2}$ that crossed the polymer and the force $F_{p,1}$ that did not.

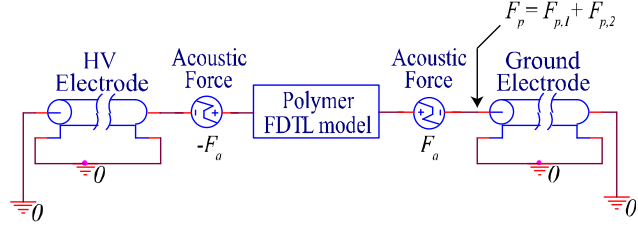

FIG. 7. Schematic of the PSpice network to show the effect of the mismatched HV electrode

FIG.8 shows plots of $F_P$ for different values of the HV electrode impedance $Z_{HV}$ compared to the polymer impedance $Z_P$. If $Z_{HV} < Z_P$, reflections between the electrode and the polymer lead to constructive interference whereas if $Z_{HV} > Z_P$ they lead to destructive interference.

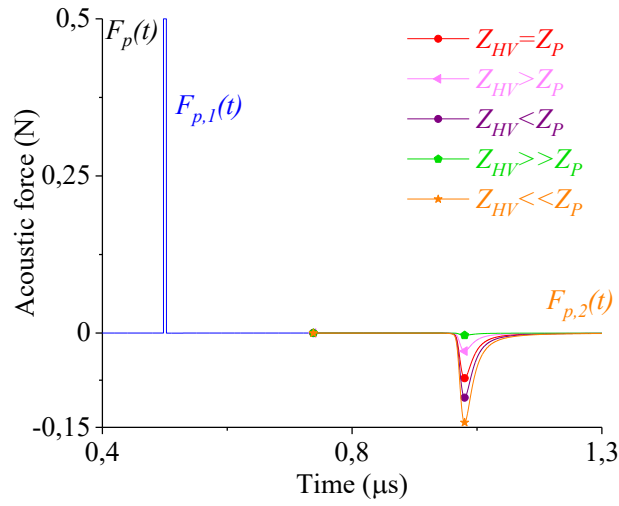

FIG. 8. Acoustic forces at each end of the polymer

The mismatching between polymer and electrodes leads to new expressions of the probed acoustic forces[41]:

$$F_{p,1}(j\omega) = G_{P \to G} F_a(j\omega) \tag{26}$$

$$F_{p,2}(j\omega) = -G_{HV \to P} T_{P \to G} F_a(j\omega) e^{-\left[\alpha_v(f) + \frac{j2\pi f}{c_{ac}(f)}\right]l} \tag{27}$$

$G_{P \to G}$ and $G_{HV \to P}$ are respectively the amplitude generation coefficients at the HV and the ground electrode interfaces. $T_{P \to G}$ is the amplitude transmission coefficient at the ground electrode interface.

$$G_{P \to G} = \frac{Z_G}{Z_P + Z_G} \tag{28}$$

$$G_{HV \to P} = \frac{Z_P}{Z_{HV} + Z_P} \tag{29}$$

$$T_{P \to G} = \frac{2Z_G}{Z_P + Z_G} \tag{30}$$

$Z_{HV}$, $Z_P$ and $Z_G$ are respectively the mechanical impedances for the HV electrode, the polymer material and the ground electrode. Applying the procedure of Ditchi *et al.*[40] to the acoustic forces, (26) and (27), provide the attenuation $\alpha_{exp}$ for the experimental configuration:

$$\alpha_{exp}(f) = -\frac{1}{l}ln\left(\left|\frac{FFT(-F_{p,2}(t))}{FFT(F_{p,1}(t))}\right|\right) = -\frac{1}{l}ln\left(\left|2G_{HV\to P}e^{-\left[\alpha_v(f)+\frac{j\omega}{c_{ac}(f)}\right]l}\right|\right) \quad (31)$$

$$\Leftrightarrow \quad \alpha_{exp}(f) = \alpha_v(f) - \frac{1}{l}ln(2G_{HV\to P}) \quad (32)$$

Note that a negative factor is applied to $F_{p,2}(t)$ to compensate the opposite sign in the attenuation and wave velocity calculations. Furthermore, the polymer attenuation is modeled as $\alpha_v(f) = a_1 f$ (cf. Equation (13)). Substitution of (13) into (32) gives:

$$\alpha_{exp}(f) = a_1 f - \frac{1}{l}ln(2G_{HV\to P}) \quad (33)$$

$$\alpha_{exp}(f) = a_1 f - \frac{1}{l}ln\left(\frac{2Z_P}{Z_{HV}+Z_P}\right) \quad (34)$$

This last equation shows that the attenuation measured from an experimental signal has an offset only due to the generation factor between the HV electrode and the polymer. Note that the change in amplitude of $F_{p,2}$ does not affect the phase difference of the wave velocity. FIG. 9 shows the experimental attenuation for different values of the HV electrode impedance.

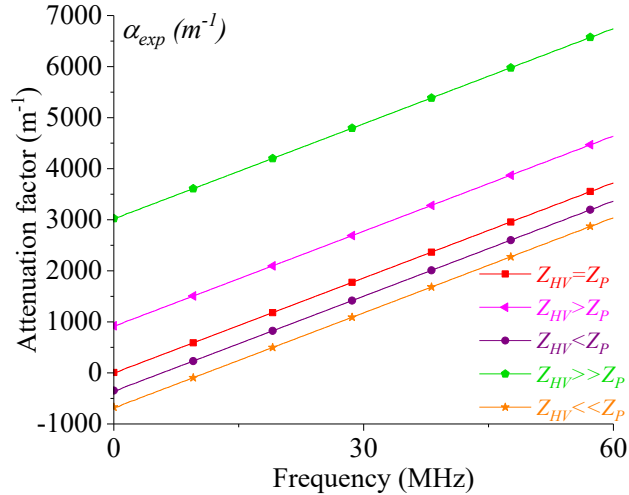

FIG. 9. Attenuation factor

This interpretation of the offset, as just related to the generation factor $G_{HV\to P}$, is a real asset to set the PEA model. Indeed, measuring the generation factor and knowing the polymer impedance provide the value of the HV impedance which is mostly not well-known.

## II. Simulation versus experiment

The PEA device has a contact area of $S = 79\ mm^2$ and is set with a PTFE sample because the wave dispersion traveling through it is easily visible even for some hundred microns. The PEA is used under biasing configuration (cf. FIG. 1) with a $1\ kV$ bias voltage and the electric pulse used is shown below:

|  | HV Electrode | Sample | Ground Electrode | Acoustic transducer | Absorber |
|---|---|---|---|---|---|
| **Materials** | Carbon black-filled LDPE | PTFE | Aluminum | PVDF | PMMA |
| **Thickness** $l\ (\mu m)$ | 1000 | 205 | 10000 | 9 | 2000 |
| **Density** $\rho\ (g/cm^3)$ | 940 | 2200 | 2700 | 1780 | 1190 |
| **Speed of sound** $c_{ac}\ (m/s)$ | 2200 | $c_{ac}(f)$ | 6400 | 2200 | 2750 |
| **Dielectric constant** $\varepsilon_r$ |  | 2.1 |  | 10 ($< 1 kHz$) |  |
| **Dielectric loss angle** $\delta_d$ |  |  |  | 0.08 ($< 1 kHz$) |  |

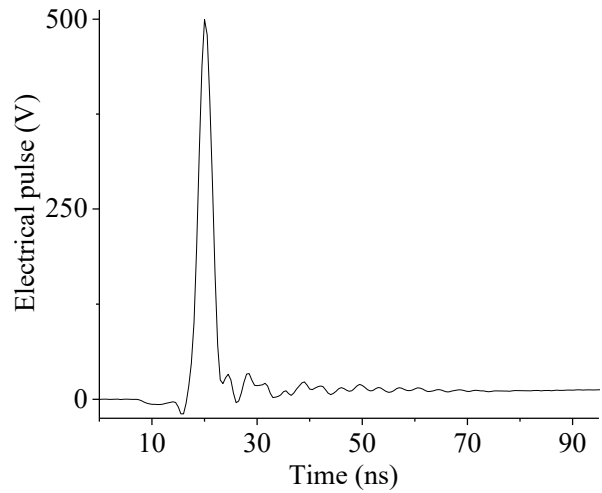

*FIG. 10. Experimental pulse*

The whole setting of the model is given in TABLE. 1 and the experimental signal in FIG. 11:

*TABLE. 1. Experimental setting*

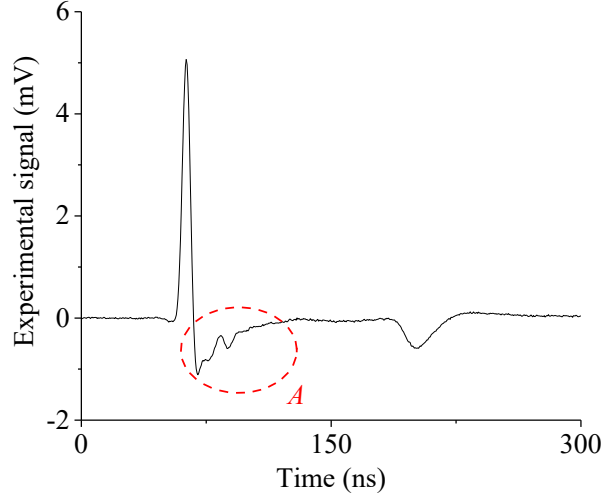

FIG. 11. Experimental PEA signal

The A overshoot is due to the presence of the amplifier as Chahal[17] showed. Under the assumption that the frequency response of the transducer coupled to the amplifier is constant in the frequency range of interest, then its transient output voltage is proportional to its transient input pressure force. This leads to apply the method of Ditchy et al[40] right to the PEA experimental signal with the aim to plot experimental attenuation and velocity as shown in FIG. 12.

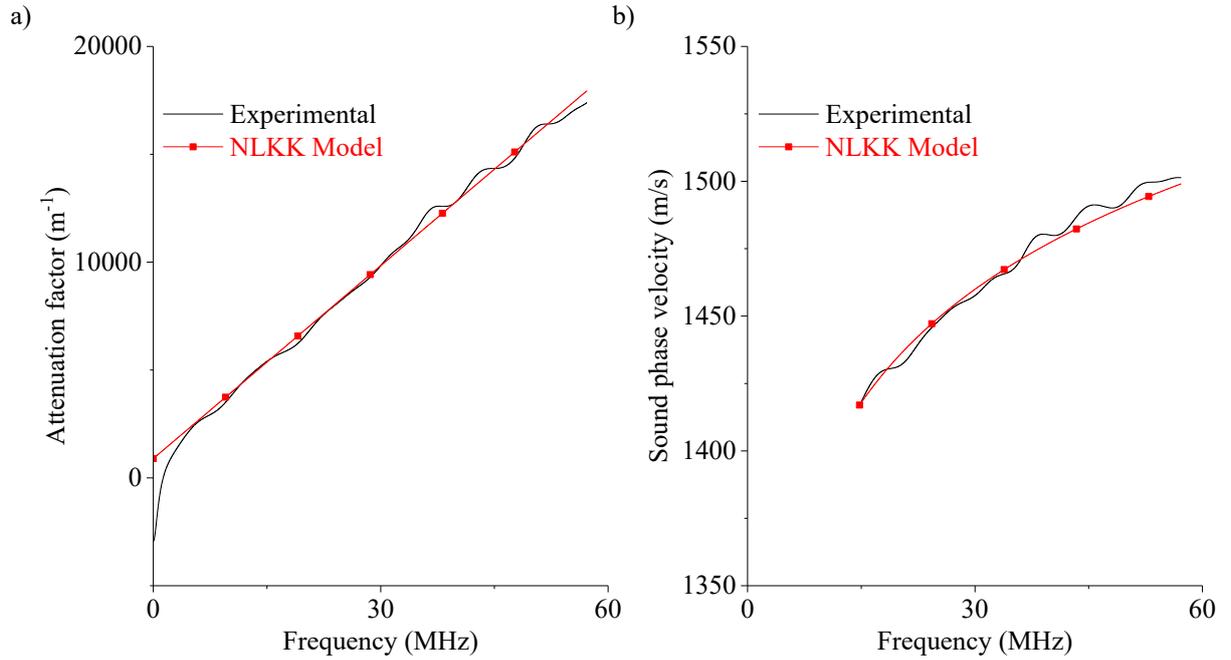

FIG. 12. (a) Attenuation factor, (b) Wave velocity

The workable bandwidth of the PEA is $60\ MHz$. Applying the regressions from NLKK model equations (34) and (14) to the experimental plots of FIG. 12.a and FIG. 12.b, gives an estimation of the polymer acoustic parameters: $Z_{HV}$, $a_1$, $f_0$ and $c_{ac}(f_0)$:

$$Z_{HV} = 294\ N.s/m$$

$$a_1 = 2{,}98.10^{-4}\ (m.Hz)^{-1}$$

$$f_0 = 14.8\ MHz$$

$$c_{ac}(f_0) = 1417 \; m/s$$

The polymer sample and then the whole PSpice model is set now. A comparison between a simulation with the standard setting and the experiment is plotted in FIG. 13. Note that the simulated signal is filtered by a Gaussian function to respect the PEA bandwidth. Our standard model bandwidth is higher than the experimental one. That could be due to some parameters not modeled in our simulation such as the aluminum ground electrode dispersion. Note that, the simulated signal amplitude has been normalized to the experimental one and will always be for the remaining graphs. Indeed, the maximum amplitude of the simulation signal is supposed to be higher. This difference may be explained by the non-modelling of the tension divider bridge inside the exciting head and between the pulse generator and an aluminum affixed to the semi-conducting HV electrode. This amplitude difference can also be explained by an inaccurate modeling of parameter such as the electromechanical piezoelectric ratio, as well as not considering the amplifier bandwidth, the oscilloscope bandwidth, and the ground attenuation.

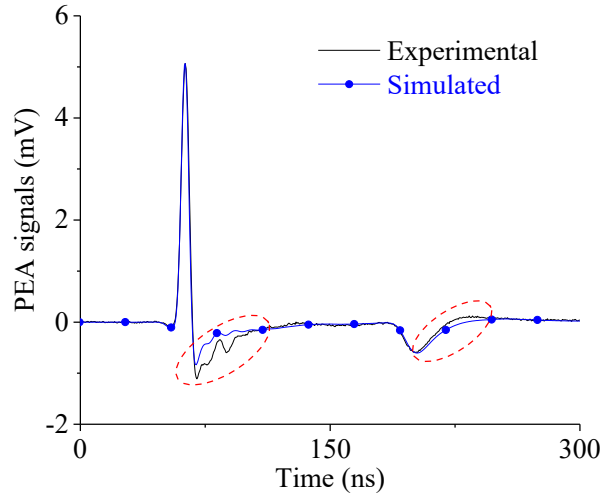

*FIG. 13. Superposition of the experimental signal and the simulated signal, which is filtered and normalized*

The simulation is still not accurate enough, especially for the circled parts that are downstream the voltage peaks that can be due to the high sensitivity of some of the model parameters. It means that, compared to other parameters, a small deviation of these parameters can significantly impact the results of the simulation. For example, the acoustic sensor length is a sensitive parameter since it can easily change the bandwidth of the simulation and the position of the A overshoot (cf. FIG. 11). An optimization procedure is applied to the sensitive parameters vector $\theta$ to minimize the following quantity:

$$S(\theta) = \left\| V_{sim}(\theta, t) - V_{exp}(t) \right\|_2^2 \qquad (35)$$

The minimization is achieved by an iterative method that adjust, step-by-step, the values of $\theta$ until the quantity $S(\theta)$ is low enough, as fixed by the tolerance threshold: $S(\theta) < \delta_S$.

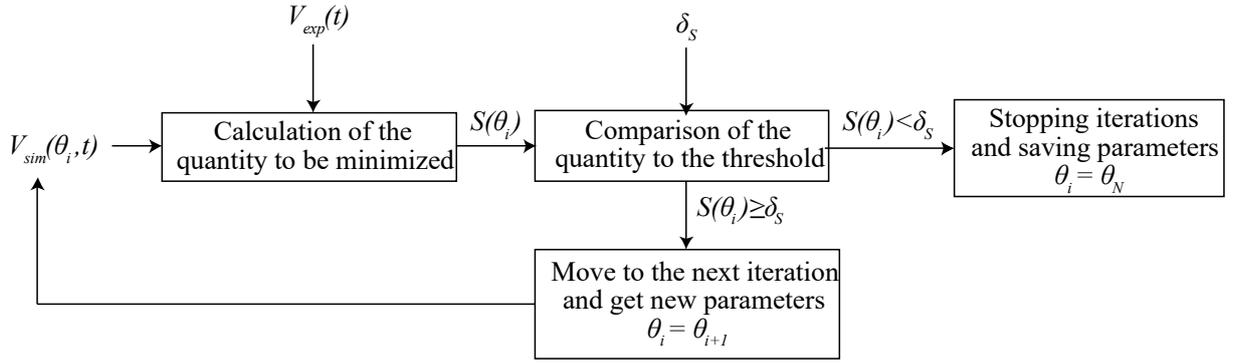

*FIG. 14. Schematic of the optimization loop*

The result of the optimized setting is shown in TABLE. 2:

*TABLE. 2. Setting of the parameters vector before and after the optimization*

|  | Before optimization | After optimization |
|---|---|---|
| **Standard deviation of the Gaussian filter ($ns$)** | 1.41 | 1.39 |
| **Sensor thickness ($\mu m$)** | 9 | 10.9 |
| **Sensor velocity ($m/s$)** | 2200 | 2250 |
| **Sensor mechanical impedance ($N.s/m$)** | 308 | 325 |
| **Sensor dielectric constant** | 10 | 7 |
| **Sensor dielectric loss angle** | 0.08 | 0.02 |
| **Absorber mechanical impedance ($N.s/m$)** | 257 | 295 |

Most of the optimized parameters have less than 5 % relative error in respect to their initial values. The 21 % relative difference for the sensor thickness can be explained by the low resolution of the length measurement device used. The 15 % relative difference for the absorber impedance can be explained by the lack of knowledge of the exact nature of the PMMA absorber used. The dielectric constant and the sensor dielectric loss angle both depend on the frequency but are modeled as constants. However, the optimized values are within the range of the frequency dependent values given in the literature[42]. The result of the new simulation with the optimized parameters is shown in FIG. 15:

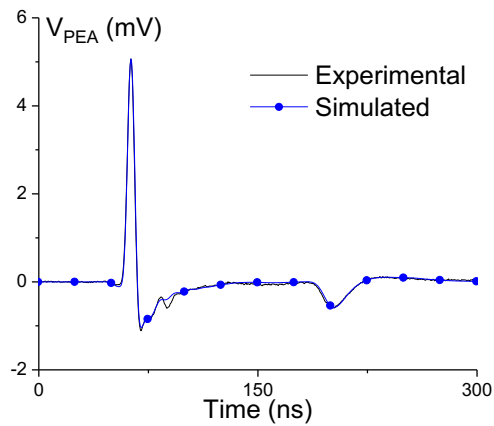

*FIG. 15. Superposition of the experimental signal and the simulated signal, which is filtered and normalized*

The percentage ratio between the norm of the difference of these above signals and the norm of the experimental signal is 7,6 %. It highlights the capacity of our model to fit the experiment.

## Conclusion

A PEA model under the PSpice software has been developed for dispersive polymer sample application. This model is an evolution regarding the previous ones thanks to the implementation of a FDTL model of the polymer sample, which takes into account acoustic wave attenuation, dispersion and reflections. Furthermore, this model provides a new method to measure, from an experimental signal, the generation coefficient at the interface between the HV electrode and the sample.
However, this evolved model will help in future studies to generate PEA transfer matrix taking into account the generation and the propagation of the pressure at different space positions inside the sample. In this way, the objectives are to use this "PSpice transfer matrix" to deconvoluate PEA signal from multi-layers or thin polymer samples.

## AIP PUBLISHING DATA SHARING POLICY

The data that support the findings of this study are available from Aurélien Pujol upon reasonable request.